\documentclass[10pt,conference,compsocconf]{IEEEtran}

\usepackage{cite}
\usepackage{amsmath,amssymb,amsfonts}
\usepackage{algorithmic}
\usepackage{graphicx}
\usepackage{textcomp}
\usepackage{xcolor}
\usepackage[caption=false]{subfig}
\usepackage{xspace}

\newcommand{\luname}{Javelin\xspace}
\newcommand{\spmv}{\texttt{spmv}\xspace}
\newcommand{\stri}{\texttt{stri}\xspace}

\begin{document}

\title{Javelin: A Scalable Implementation for Sparse Incomplete LU Factorization
\thanks{This work used the Extreme Science and Engineering Discovery Environment (XSEDE), which is supported by National Science Foundation grant number ACI-1548562.
Support for this work was also funded by Franklin \& Marshall Hackman Fund.}
}

\author{
\IEEEauthorblockN{Joshua Dennis Booth}
\IEEEauthorblockA{\textit{Department of Computer Science} \\
\textit{Franklin \& Marshall College} \\
Lancaster, USA \\
joshua.booth@fandm.edu}
\and
\IEEEauthorblockN{Gregory Bolet}
\IEEEauthorblockA{\textit{Department of Computer Science} \\
\textit{Franklin \& Marshall College} \\
Lancaster, USA \\
gbolet@fandm.edu}
}

\maketitle

\begin{abstract}
In this work, we present a new scalable incomplete LU factorization framework called \luname to be used as a preconditioner for solving sparse linear systems with iterative methods.
\luname allows for improved parallel factorization on shared-memory many-core systems by packaging the coefficient matrix into a format that allows for high performance sparse matrix-vector multiplication and sparse triangular solves with minimal overheads.
The framework achieves these goals by using a collection of traditional permutations, point-to-point thread synchronizations, tasking, and segmented prefix scans in a conventional compressed sparse row format.
Moreover, this framework stresses the importance of co-designing dependent tasks, such as sparse factorization and triangular solves, on highly-threaded architectures.
Using these changes, traditional fill-in and drop tolerance methods can be used, while still being able to have observed speedups of up to $\sim 42 \times$ on 68 Intel Knights Landing cores and $\sim 12 \times$ on 14 Intel Haswell cores.
\end{abstract}


\section{Introduction}
\label{sec:intro}
The solution to sparse linear systems of equations dominates the execution time of most parallel scientific simulations.
A common approach to solving such systems is using iterative methods, e.g., Conjugate Gradients (CG) or Generalized Minimal Residual Method (GMRES), that rely on sparse matrix-vector multiplication (\spmv) and sparse triangular solves (\stri).
However, iterative methods may not converge or take a large number of steps depending on the conditioning of the system, and some form of preconditioning is normally used.
Incomplete factorization is one common method used to precondition the system in order to improve the iterative method by factorizing the coefficient matrix representing the linear system of equations while controlling the number of nonzero elements calculated during the factorization.
This incomplete factorization used as a preconditioner allows for the linear system to be modified to:
$Ax = b \rightarrow M^{-1}Ax = M^{-1}b.$
Because the precondition method tries to limit the number of calculations with $M$, the ratio of float-point operations required to construct $M$ and solve $M^{-1}b$ is memory-bound making incomplete factorization extremely hard to achieve strong-scaling. 
One common and useful preconditioner ($M$) is based on the incomplete LU factorization. 
In this work, we provide a new scalable incomplete LU factorization (ILU) framework called \luname that is designed for current large many-core systems, and allows for scalable sparse triangular solves without having to reformat the input matrix like most scalable triangular solve methods~\cite{sts} . 

A rich variety of incomplete LU factorization algorithms exist, such that $A \approx M = LU$ where $L$ and $U$ are lower and upper triangular matrices, respectively.  
In general, these algorithms can be categorized by the method they use to reduce nonzeros due to fill-in: dropping based on numerical value (ILU($\tau$)) and dropping based on fill-in level (ILU(k)).
The design of \luname allows for any combination of the two to be used (ILU(k,$\tau$)) along with modified ILU~\cite{milu}.
However, most of the current packages that can be used for traditional incomplete factorization are serial because of the difficulty of scaling and the overhead that may exist to reformat the output matrix in order to achieve scalable sparse triangular solves.
Though parallel ILU packages exist~\cite{fastilu,Henon2006,boothcompare,Hysom2001}, many rely on complex data-structures or modifications of ILU by using approximations to reduce synchronizations, and may not have scalable \stri. 
These packages apply these modifications due to the difficulty in achieving scalable performance with traditional ILU.
\luname is designed to have the robust nature of traditional ILU methods while limiting the use of complex data-structures and still achieving the scalability needed for incomplete factorization, \spmv, and \stri on current many-core systems such as Intel Xeon Phi by co-optimizing incomplete factorization and \stri.

This paper presents \luname as implemented currently as a threaded templated C++ incomplete LU factorization using OpenMP, and having the following contributions:
\begin{itemize}
	\item A parallel ILU optimized for parallel \spmv and \stri on many-core architectures
	\item A parallel ILU that requires minimal data preprocessing and permutations
	\item An empirical evaluation of \luname on Intel Xeon (Haswell) and Intel Xeon Phi (Knights Landing) for ILU(0) and \stri
        \item An evaluation of optional parameters 
\end{itemize}

\section{Background}
\label{sec:background}
This section provides a brief overview of the foundations and improvements of incomplete LU factorization, i.e., $A \approx M = LU$, and current scalable implementations of sparse matrix-vector multiplication (\spmv) and sparse triangular solve (\stri).
Here, \spmv and \stri methods are the driving force behind this work as we desire to have a scalable incomplete factorization that is optimized to the state-of-the-art parallel many-core \spmv and \stri as they dominate the time of iterative methods.

\textbf{Incomplete LU factorization.}
Preconditioning based on incomplete LU decomposition is generally regarded as a robust ``black-box" method for unstructured systems arising from a wide range of application areas.
ILU has been studied extensively.  
This includes extensive studies on variations of ILU(k), ILU($\tau$), and multiple levels of these two together~\cite{milu,Chow1997}.
The choice of which combination is highly dependent on the linear system.
Variations exist that try to improve serial performance using supernodes~\cite{superlu}.
However, unstructured systems with limited fill-in will have limited amount of overlapping sparsity patterns for supernodes.

\textbf{Parallel incomplete LU factorization.}
The first work on parallel incomplete factorization was for regular structured systems.
However, the need for parallel incomplete factorization on unstructured systems resulted in much work throughout the 1980s until today.
The traditional approach in dealing with this irregular problem of unstructured systems is to use \textit{level scheduling}.
In level scheduling, the coefficient matrix is transformed into a graph, such that rows, columns, or blocks of the coefficient matrix represent vertices and dependencies are represented as edges in a graph.
Groups of vertices in the graph that do not have any incoming edges are placed together into a level, and the outgoing edges from these vertices are removed.
The rows, columns, or blocks that are in each level can be scheduled to be solved concurrently. 
In a similar fashion, orderings, such as Coloring and Nested-Dissection, have been used to identify rows, columns, or blocks that can be factorized concurrently while in some cases exposing more concurrency.
The disadvantage to all of these reorderings is that they \emph{may} negatively impact the number of iterations needed for the iterative method to converge~\cite{order1}.
We explore this possible negative impact with multiple reordering in section~\ref{sec:anal}.
However, the performance of parallel incomplete factorization depends on many more factors than the exposure of the levels of parallelism.
These factors include programming model, task size, and data-structure related to the targeted machine.
This was first noted in the early 1990s~\cite{cgmem}, and recently examined on current many-core systems with different factorization codes in~\cite{boothcompare}.
Recently, one version of ILU has been shown to achieve very good performance on many-core and GPU systems~\cite{fastilu}.
Despite performance, this method may result in an incomplete factorization that is nondeterministic and that challenges traditional dropping or modified incomplete factorization due to race conditions.

\textbf{Sparse matrix-vector multiplication and triangular solve.}
Iterative methods, such as CG and GMRES, spend the vast majority of their time preforming the operations of sparse matrix-vector multiplication (\spmv) and sparse triangular solve (\stri). 
In particular, preconditioned CG using incomplete Cholesky Decomposition, i.e., $M = LL^T$, spends up to $70\%$ of its execution time in forward and backward \stri~\cite{park}.
Because these operations dominate the solve time, having an incomplete decomposition that can allow us to make use of these operations efficiently is critically important. 
There has been a vast amount of research in this area over the past 10 years as large many-core systems are becoming the norm~\cite{park,andrew,sts,csr5}.
Many of these methods require reordering and special data-structures~\cite{sts,csr5}.
Therefore, these methods are not ideal for iterative methods as matrices would need to be copied over to these data-structures, and the number of iterations would depend on ordering.
Moreover, these orderings may not be ideal for parallel incomplete factorization.
Two works standout that require us to limit reordering and data movement.
The first work~\cite{csr5} introduces an efficient storage format for \spmv, namely CSR5.
CSR5 does not require a reordering of sparsity pattern, depends on the common Compressed Sparse Row (CSR) format, and allows for quick segmented scans.
The only overhead needed is a little extra storage for tile information to help support segmented scan operations.
Segmented scan has been shown to be an ideal way to implement \spmv and \stri on vector based machines~\cite{mikeseg}, with most many-core systems having large vector register lanes.
The second work~\cite{park,andrew} focuses on improving \stri on standard CSR format storage using general level scheduling and sparsifying synchronization.
This work demonstrates that many of the overheads dealing with level scheduling could be removed, and allowing \stri to scale.
\luname uses these works as the foundation to achieve scalable \stri, with modification that allow incomplete factorization. 

\vspace{-6pt}
\section{A framework for  ILU}
\label{sec:alg}
The \luname framework for parallel incomplete LU factorization depends on predetermining the sparsity pattern and applying an up-looking LU algorithm, i.e., a left-looking algorithm done by rows instead of columns, to the pattern.  
Both of these operations need to be done in parallel for the factorization code to scale on modern many-core systems.
Determining the sparsity pattern in parallel has been studied in the following work~\cite{Hysom2001}, and a good implementation exists in~\cite{tacho,boothcompare}.
Therefore, we will primarly focus on the second half of implementing the up-looking LU algorithm to the given nonzero pattern.
Fig.~\ref{alg:lu} provides an example of the up-looking LU algorithm using standard MATLAB notation for the sparse matrix $S$ containing the sparsity pattern, which is used by many incomplete factorization codes.
The main reason for the popularity of up-looking LU is because it can be easily modified to be used with dropping based on some tolerance to the pivot value and modified ILU techniques which help to reduce iteration counts~\cite{milu}.
Moreover, up-looking LU allows for local estimates of resilience from soft-errors and the convergence rate.
Like many other incomplete factorizations, \luname does not allow for pivoting.
These algorithms could be applied to other preconditioners and is why we call \luname a framework.
\vspace{-5pt}
\begin{figure}[tbh]
\centering

\scalebox{.90}{
\begin{minipage}{0.5\textwidth}

\begin{algorithmic}[1]
\REQUIRE $S$ contains the sparsity pattern for $L$ and $U$, and $L$ and $U$ are stored in $A$
\FOR{$row = 1$ to $n$ }
	\STATE $s_{row} = S(row,:)$
	\FOR{$col$  in $s_{row}$}
		\IF{ $col < row$}
			\STATE $a_{row,col} = a_{row,col}/a_{col,col}$
			\STATE $s_{update}  = S(col,:)$
			\FOR{$update\_col$ in $s_{update}$} 
				\IF{$update\_col$ in $S(row,:)$ and $update\_col > col$}
					\STATE $a_{row,update\_col}$ -= $a_{row,col} \times a_{col, update\_col}$ 
				\ENDIF 
			\ENDFOR
		\ENDIF
	\ENDFOR
\ENDFOR
\end{algorithmic}

\end{minipage}
}

\caption{Up-looking Sparse Incomplete LU Factorization Algorithm}
\label{alg:lu}
\end{figure}

\vspace{-6pt}
In addition, \luname uses level scheduling as the primary method to apply up-looking LU.
Past work related to level scheduling of \stri shows it is difficult to achieve high performance, with applications scaling to only a few cores.
These difficulties were due to the requirement of thread barriers between levels and levels containing too few rows to keep all threads busy.
To the best of our knowledge, this is the first time using level scheduling for up-looking LU using \stri access pattern. 
\luname faces these issues head-on by using point-to-point synchronizations between levels, and then using a second stage approach when the size of levels becomes too small for the thread count.
Two different methods exist for factoring the second stage, namely Segmented-Row and Even-Row.  
We will explain both along with their trade offs, and demonstrate their performance differences in \luname, though \luname by default will make the choice for the user based on the matrix structure.
In order to apply this two-stage structure, we first find the level scheduling order to either $lower(A)$ or $lower(A+A^{T})$ depending on the user options, where $lower$ is the sparsity pattern of the lower triangular half of the given matrix.
We address the importance of the option between $A$ and $A+A^{T}$ in the subsection related to the second stage factorization method.
Though finding the order alone would be enough to schedule the top stage point-to-point factorization method, we permute the nonzeros in the matrix into the level ordering while copying $A$ into the CSR data-structure of $L$ and $U$ in parallel allowing for first-touch. 

\begin{figure}[tbh]
	\centering
	\includegraphics[scale=0.28]{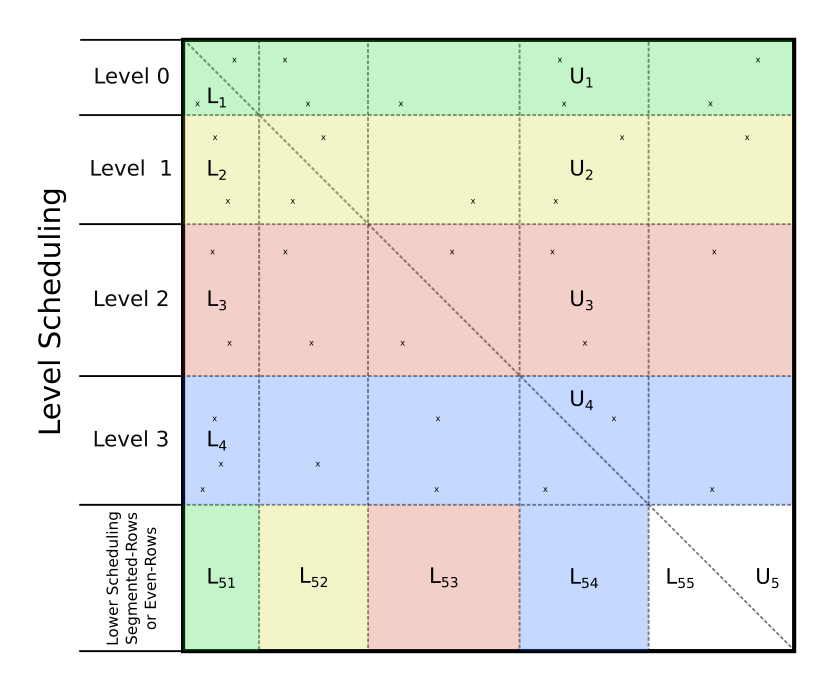}
	\caption{The general structure generated by level scheduling ordering.  Levels in the top half can be factored in an up-looking method while point-to-point synchronizations are used to handle dependencies.  The last level can be solved using one of two different methods.}
	\label{fig:struct}
\end{figure}

\vspace{-6pt}
\subsection{Level Scheduling, Upper Stage}
\begin{figure}[thb]
	\centering
	\includegraphics[scale=0.28]{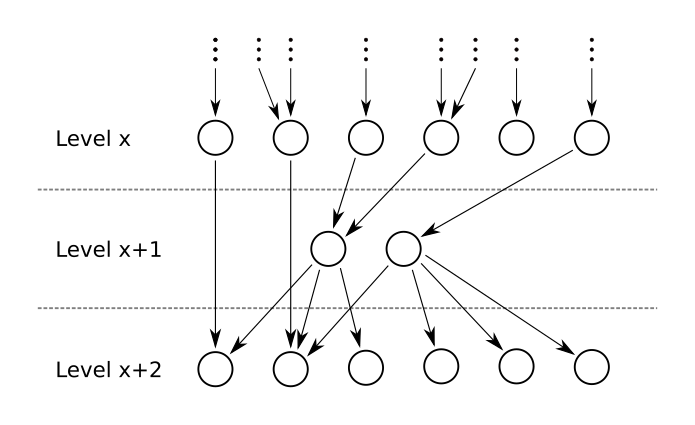}
	\caption{Here we have a large number of tasks in level x and x+2.  However, between them is a level with few tasks. Therefore, this level may not be ideal to be solved by our second level method.}
	\label{fig:bsb}
\end{figure}

During preprocessing, the sparsity pattern of $A$ with desired fill-in is found and permuted into a level-based ordering as in Fig.~\ref{fig:struct}.
\luname uses a set of user defined options to determine the portion of the matrix to factor using level scheduling.
These options include minimal number of rows per level, relative location of level in ordering, and row density, i.e., the number of nonzeros per row.
The idea is that \luname will want to apply a level scheduling to levels that have a very large number of rows so that no thread will run out of work and any imbalance in the amount of work would be amortized across multiple rows.
However, if the rows become too dense compared to the relative average density of the matrix or the number of rows in the level is too few, these are moved towards the end to be solved with \luname's second stage solver.
The catch is when there exists a level with very few rows or high row density in the middle of larger level sets, see Fig.~\ref{fig:bsb}. 
In such cases, moving these rows to the second stage would result in moving a lot of work to the second stage (i.e., all the dependent rows).
Traditional implementations of level scheduling would use barriers and would have difficulty because of such rows.
However, \luname using point-to-point synchronizations does not suffer in the same way.

\begin{figure}[thb]
	\centering
	\includegraphics[scale=0.28]{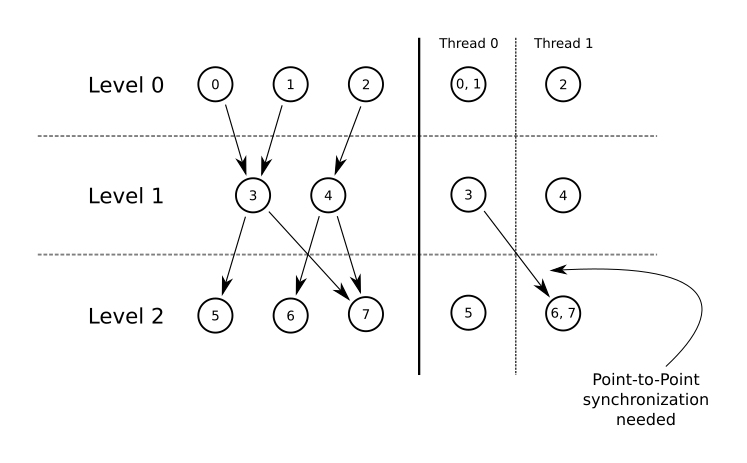}
	\caption{In the left side of the figure, the tasks and their dependencies that are generated by level scheduling.  In the right side of the figure, tasks map to threads while imposing an implied ordering of tasks in a level.  Using this order, dependencies are pruned. }
	\label{fig:p2p}
\end{figure}

\vspace{-8pt}
Point-to-point synchronizations for level scheduling was introduced as an effective way to parallelize sparse triangular solve~\cite{park,andrew}.
Here we use the observation that incomplete factorization done with up-looking LU has the same dependency structure as sparse triangular solve.
The principle of sparsifying point-to-point synchronizations for level scheduling is as follows.
The level sets are found as in Fig.~\ref{fig:p2p}.
Once the sets are found, rows are mapped to threads.
This mapping of rows to threads induces an implied ordering, and the implied ordering is used to prune the full set of dependencies.
 In traditional methods, these dependencies are taken care of by a global barrier between levels or by generating tasks with dependencies in a parallel tasking programming model.
 Both of these methods have high overheads at runtime.
 However, point-to-point's implementation relies on inexpensive spinlocks and allows for certain threads to speed ahead of others.
 
 \vspace{-8pt}
\subsection{Lower Stage}
At some point, level scheduling will no longer offer the degree of concurrency needed to achieve continued speedup on modern many-core systems.
In these cases, \luname will switch to one of two different methods.
We name these methods: Segmented-Rows and Even-Rows.

\begin{figure}[th]
        \centering
        \includegraphics[scale=0.34]{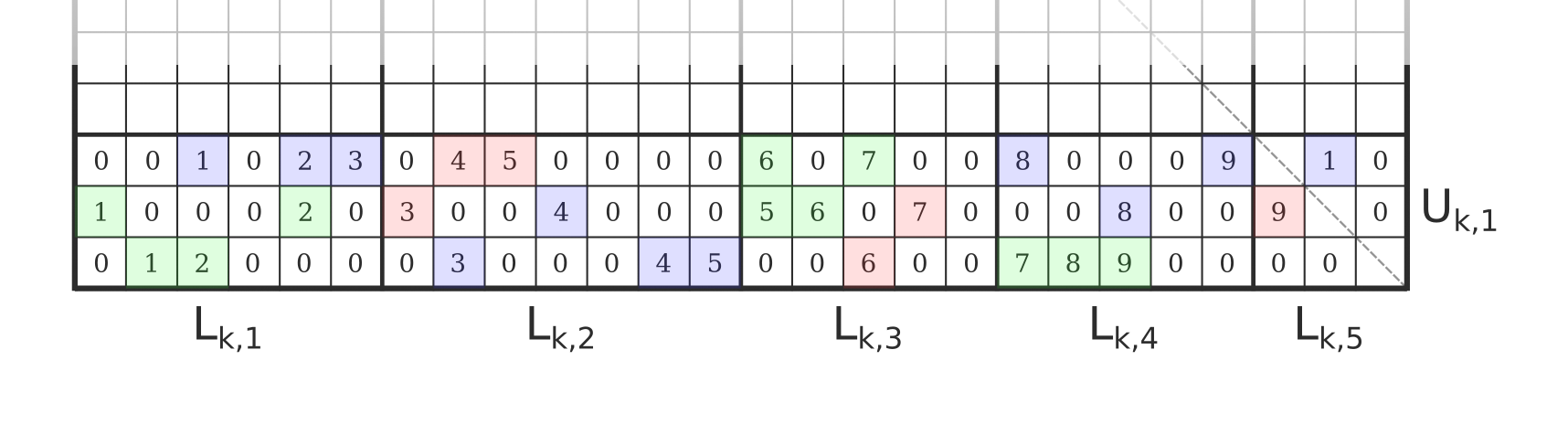}
        \caption{Layout of blocks and tiles in SR method.  Titles are laid out in each block ($L_{k,i}/U_{k,1}$) independently of each other, and can span multiple rows. Nonzero elements in each row are indexed starting from 1. Contiguous nonzeros that are grouped together are colored similarly.}
        \label{fig:sr}
\end{figure}

\textbf{Segmented-Rows.}
The Segmented-Rows (SR) method is inspired by the segmented scan that achieves cross-platform scalability of \spmv in CSR5~\cite{csr5}.
The goal is to have a structure that can be factored efficiently while already being setup for a \spmv -like update needed by \stri. 
In preprocessing, the level ordering of $lower(A+A^{T})$ is found, and the rows identified to have SR applied to them are permuted to the end of the matrix during the copy-fill-in phase of building a CSR data-structure.
Fig.~\ref{fig:sr} provides a visual example of this format, where the index of each nonzero element in a row is numbered starting from 1.
Each subblock ($L_{k,i}$) is defined by the level scheduling using in the upper stage.
Inside each subblock, nonzeros are grouped together in a contiguous manner, as displayed in Fig.~\ref{fig:sr} by different colors.
Given a subblock ($L_{k,i}$) formed based on the level scheduling ordering, we note that there are no dependencies between the columns within the same subblock, i.e., $L_{k,i}(:,x)$ does not depend on $L_{k,i}(:,y)$ where $y > x$. 
This fact is due to using the level ordering for $lower(A+A^{T})$ in place of $lower(A)$, which the latter does not guarantee the above observation. 
This observation does allow for tiles to be created within each level of the subblock in a similar manner to CSR5 that only requires a small additional array of pointers into the block.
These tiles build the foundation to the SR algorithm.
A tasking model is used to apply factorization to each level, then to apply the needed update to other tiles in $L$ and $U$, and allowing the next subblock's set of tiles to be factored. 
In \luname, OpenMP tasking is used as the programming model, and tile size options are made available to the user.
Fig.~\ref{alg:segrow} provides the algorithm used, and for simplicity we do not explicitly state the dependencies.

\begin{figure}[tbh]
\centering

\scalebox{0.9}{
\begin{minipage}{.5\textwidth}

\begin{algorithmic}
\REQUIRE Let $k$ be the level to complete SR.
	\FOR{$i = 0$ to $num\_lvl$}
		\STATE spawn all tiles (DIVIDE\_COLUMNS($L_{k,i}$))
		
		\FOR{$j=i$ to $num\_lvl$}
			\STATE \textbf{spawn} all tiles (UPDATE\_BLOCK($L_{k,i}$, $L_{k,j}$))			
			
		\ENDFOR 
	\ENDFOR
	\STATE \textbf{spawn} all tiles (FACTOR\_LU())
\end{algorithmic}

\end{minipage}
}

\caption{Segmented-Row Method (SR) Algorithm}
\label{alg:segrow}
\end{figure}

\vspace{-4pt}
In Fig.~\ref{alg:segrow}, \texttt{DIVIDE\_COLUMNS} takes the diagonal value in $U$ corresponding to the used columns in the tile and divides the entries in the tile.
The \texttt{UPDATE\_BLOCK} method applies the multiplication-subtraction updates to nonzeros in tiles in later levels, similar to the inner-most line in Fig.~\ref{alg:lu}.
Lastly, \texttt{FACTOR\_LU} factors the tiles in the blocks $L_{k,k}$ and $U_{k,1}$. 

SR is useful in several different situations.  
These situations are when there exist fewer rows that are excluded from the level scheduling method than threads running on the system and the number of nonzeros in the rows are highly imbalanced.
However, SR suffers from the need to use a tasking programming model and the overheads that come with it.
Moreover, the fixed tile size of SR lends itself to using optimized vector operations (e.g., AVX).
Many new hardware such as Intel Xeon Phi have large vector lanes that need to be used to achieve top performance.

\begin{figure}[th]
        \centering
        \includegraphics[scale=0.30]{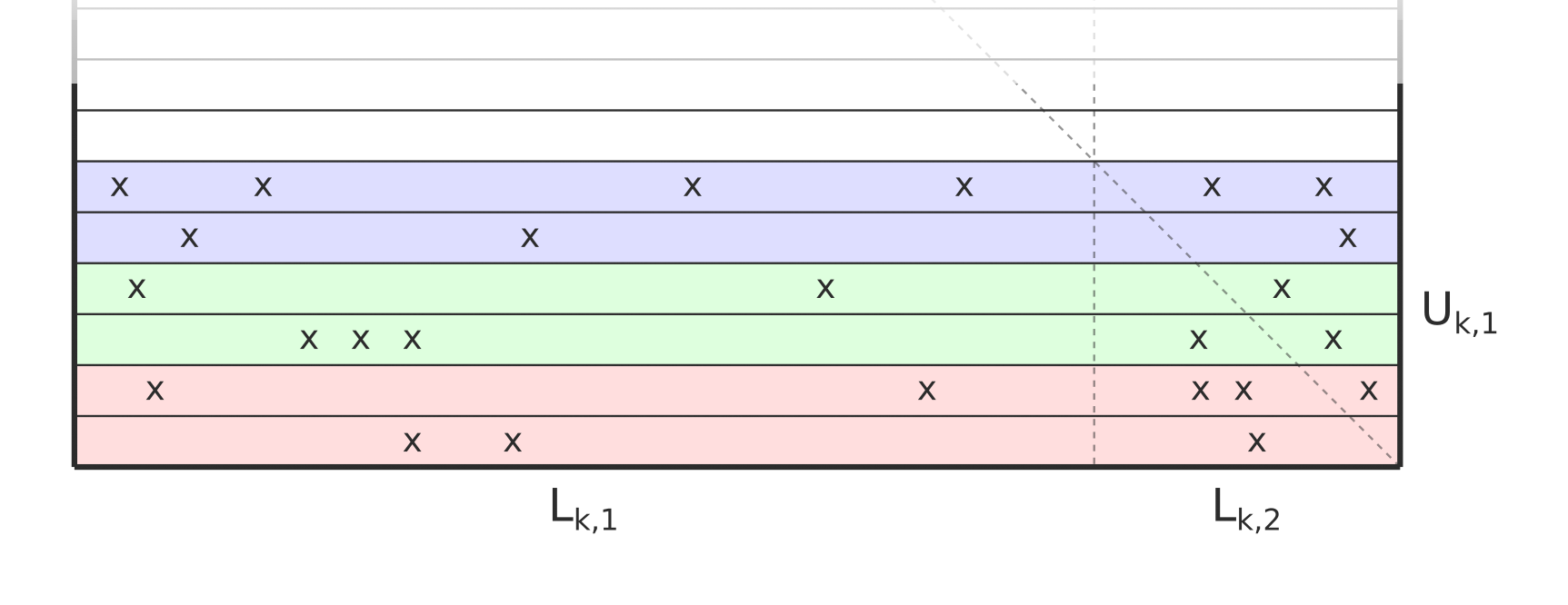}
        \caption{Layout of blocks in the ER method.  Each thread will get a number of rows.  They will complete an up-looking incomplete LU on $L_{k,1}$ while making the needed updates to the values in $L_{k,2}$ and $U_{k,1}$.}
        \label{fig:er}
\end{figure}

\textbf{Even-Rows.}
The Even-Rows (ER) method is a much simpler method than SR.
ER depends on the number of rows excluded from the level scheduling method being greater than the number of desired threads to be used on the system.
With the number of rows being greater than the number of threads, each thread can factor multiple rows and any imbalance can be averaged over all thread's rows.
The excluded rows are reordered to the end of the matrix similar to SR after finding the level order. 
Here, either $lower(A+A^T)$ or $lower(A)$ could be used to find the level order.
While $lower(A)$ generally offers more and larger levels, this does not seem to have much difference in performance, see section~\ref{sec:anal}.
Options exist to allow in preprocessing to form tiles similar to SR to be added for \stri, but are not needed for the incomplete factorization.
In Fig.~\ref{fig:er}, each thread gets a row and factors $L$ up until the corner piece (i.e., $L_{k,2}$ and $U_{k,1}$) while making the needed updates to the corner piece's $L_{k,2}$ and $U_{k,1}$.
Fig.~\ref{alg:er} provides an overview of the ER method, where \texttt{FACTOR\_L} does the factorization of each row up until the corner piece.
When all the threads are done with their rows, the corner piece can be factored with a call to \texttt{FACTOR\_LU}.  
The factorization of the corner can be done in serial or parallel.  
However, for most matrices, serial seems to be good enough.

\begin{figure}[tbh]
\centering
\scalebox{0.9}{
\begin{minipage}{0.5\textwidth}

\begin{algorithmic}
\REQUIRE Let $k$ be the level to complete ER.
	\FOR{$row$ in $lower$ in parallel}
		\STATE FACTOR\_L($row$)
	\ENDFOR
	\STATE \textbf{spawn} all titles (FACTOR\_LU($L_{k,2}$, $U_{k,1}$))
\end{algorithmic}

\end{minipage}
}

\caption{Even-Rows Method (ER) Algorithm}
\label{alg:er}
\end{figure}

\vspace{-8pt}
\section{Experimental setup}
\label{sec:setup}
Here we provide the setup for the next three sections to evaluate the performance of incomplete LU factorization, sparse triangular solve, and the sensitivity of \luname to input.

\textbf{Test systems.}
We test our framework on two systems, i.e., Bridges at PSC and Stampede2 at TACC.
The first system contains two Intel Xeon E5-2695 v3 (Haswell) processors each with 14 cores and having 128GB DDR4.
The second system contains an Intel Xeon Phi Knights Landing 7250 (KNL) each with 68 cores and having 96GB DDR4.
The KNL is setup in cache mode with 32KB L1, 1MB L2 per two-core tile, and 16GB direct-mapped L3.
Though KNL has a variety of different ways to setup its fast memory, such as cache mode and flat, we test in cache mode as this is the mode recommended by TACC to most MPI+X scientific applications.
Therefore, we wanted to test in the mode likely to be used by applications that will use \luname.
All codes are compiled with Intel Compiler 17.4 / 17.3 with -O3 options.
We use OpenMP with the \texttt{DYNAMIC} scheduling and \texttt{CHUNK\_SIZE=1} for all our tests, though ER may benefit from different scheduling and chunk size options.
This decision was made to limit the number of possible combinations in finding the best overall performance.

\begin{table}[tb]
\centering
\caption{Test Suite.  N is the matrix dimension, Nnz is the number of nonzero, RD is nonzeros / N, SP is if the symbolic pattern of the matrix in natural order is symmetric, and Lvl is the number of levels found in the level scheduling.}
\label{tab:matrices}{\footnotesize
\begin{tabular}{ | c | r | r | r | c | r | c | }
\hline
\textbf{Matrix} & \textbf{N} & \textbf{Nnz} & \textbf{RD} & \textbf{SP} & \textbf{Lvl} & \textbf{}  \\ \hline
wang3 & 26064 & 177168 & 6.8 & yes & 10 & B \\ \hline
TSOPF\_RS\_b300\_c2 & 28338 & 2943887 & 103.88 & no & 180 & B \\ \hline
3D\_28984\_Tetra & 28984 & 285092 & 9.84 & no & 34 & B \\ \hline
ibm\_matrix\_2 & 51448 & 537038 & 10.44 & no & 29 & B \\ \hline
fem\_filter & 74062 & 1731206 & 23.38 & yes & 554 & B \\ \hline
trans4 & 116835 & 749800 & 6.42 & no & 20 & B \\ \hline
scircuit & 170998 & 958936 & 5.61 & yes & 34 & B \\ \hline
transient & 178866 & 961368 & 5.37 & yes & 16 & B \\ \hline
offshore & 259789 & 4242673 & 16.33 & yes & 74 & A \\ \hline
ASIC\_320ks & 321671 & 1316085 & 4.09 & yes & 16 & B \\ \hline
af\_shell3 & 504855 & 1.756e7 & 34.79 & yes & 630 & A \\ \hline
parabolic\_fem & 525825 & 3674625 & 6.99 & yes & 28 & A \\ \hline
ASIC\_680ks & 682712 & 1693767 & 2.48 & yes & 21 & B \\ \hline
apache2 & 715176 & 4817870 & 6.74 & yes & 13 & A \\ \hline
tmt\_sym & 726713 & 5080961 & 6.99 & yes & 28 & B \\ \hline
ecology2 & 999999 & 4995991 & 5 & yes & 13 & A \\ \hline
thermal2 & 1.2e6  & 8580313 & 6.99 & yes & 27 & A \\ \hline
G3\_circuit & 1.5e6 & 7660826 & 4.83 & yes & 13 & B \\ \hline
\end{tabular}
}
\end{table}

\textbf{Matrices.}
We select a large array of different matrices from the SuiteSparse collection~\cite{matrix}. 
Table~\ref{tab:matrices} provides the details for each.
Our test suite of matrices is broken into two pieces (labeled A and B).
The first group (A) are commonly tested for ILU with regard to convergence~\cite{fastilu,superlu}, despite being symmetric positive definite.
The goal of group A is to demonstrate the effect our structure and preordering described in this section have on the convergence.
Though previous studies have looked at the effect of preordering on convergence, no study has looked at our particular ordering combination of level set.
This is presented in section~\ref{sec:anal}.
The second group (B) contains a wide array of matrices from multiple areas.
While much previous work
focuses on only providing empirical evaluation of matrices from areas coming from the discretization of partial differential equations (PDEs), there is a growing need for iterative methods in other areas that have very irregular matrices, such as certain stages of circuit simulation~\cite{Schilders2002}. 
Therefore, to have a thorough examination, we include a wide range of matrices with different numeric patterns and row densities, i.e., average number of nonzeros per row.

\textbf{Incomplete levels.}
We will only consider ILU($k$) with $k=0$, though \luname supports other levels as implemented by other work~\cite{Hysom2001,tacho,boothcompare} and commonly used in iterative solvers.
As $k$ increases, additional fill-in is allowed into the sparsity pattern.
The placement and amount of this would depend on the ordering of the matrix.
Therefore, comparing performance as $k$ varies would not provide a deep understanding of the scalability of \luname without knowing where and how much fill-in was produced.
It is therefore common to compare scalability primarily with ILU(0)~\cite{boothcompare,fastilu} over a large test suite of matrices with different sparsity pattern and row density in order to estimate how well the implementation scales with fill-in as we do in this paper.

\textbf{Preordering.}
It is common to permute the nonzeros of a sparse matrix before applying an iterative method.
These permutations allow for better memory access patterns and reductions to iteration counts~\cite{order1}.
\luname currently only has a parallel level-set ordering built-in that is needed for the framework. 
Therefore, we assume that the given matrix is already ordered in some manner that is ideal for the given solver method by the user.
For the performance evaluation, we assume that the given matrix has been ordered in the following manner.
A Dulmage-Mendelsohn ordering is used to move nonzeros to the diagonal of the matrix,  then Nested-Dissection (ND) from METIS is performed.
This ordering was chosen because ND is commonly applied to coefficient matrices for parallel factorization~\cite{tacho}.
We consider the sensitivity of this choice in section~\ref{sec:anal}, and compare against the use of ordering with Reverse Cuthill McKee (RCM) ordering to using ND.

\vspace{-4pt}
\section{Evaluation of Incomplete LU}
\label{sec:perf}
\begin{figure}[tbh]
\centering
\subfloat[Slowdown on Haswell]{\label{fig:hwsmp}\includegraphics[scale=0.28]{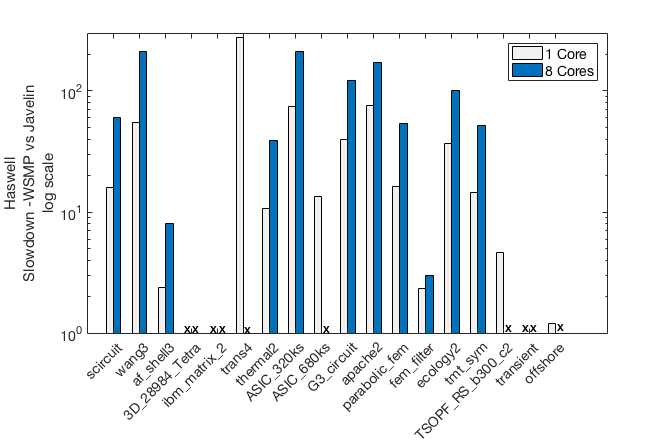}}
\\
\subfloat[Slowdown on KNL]{\label{fig:kwsmp}\includegraphics[scale=0.28]{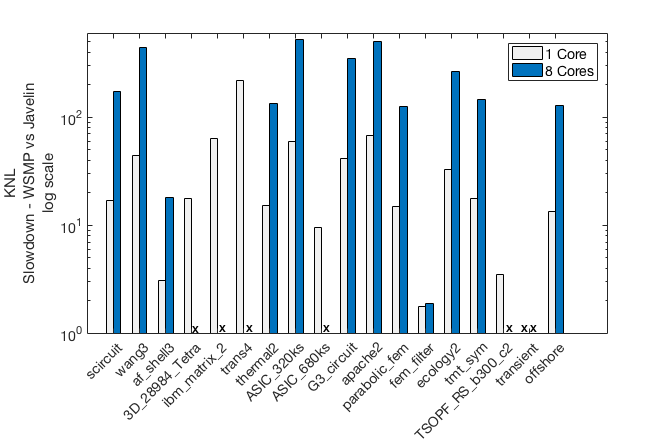}}
\caption{Slowdown of WSMP compared to Javelin on Haswell and KNL.  WSMP does not have any real scaling behavior after 8 cores. 'x' represents where WSMP failed due to internal issues.}
\label{fig:slowdown}
\end{figure}

We first considered the performance of \luname to other packages.
We note that there does not exist many threaded incomplete factorization packages.
Many popular packages (even ones in distributed memory), such as SuperLU~\cite{superlu} and Trilinos, only offer a serial option, and we found that \luname in serial was either faster than (15 matrices) or within $10\%$ (3 matrices) of these serial packages.
Additionally most packages that offer traditional ILU use only ILU(k, $\tau$) and does not provide an interface to ILU(k).
Therefore, we will use this double method to compare to the commercial package Watson Sparse Matrix Package (WSMP)~\cite{wsmp2}, but will use ILU(k) for scalable performance analysis for reasons stated before.
WSMP is given the matrix in the level ordering used by \luname with k=0 and $\tau$ is set to be a value so that nonzeros are similar to that of  ILU(0), and  we do not allow pivoting.
Here, only \luname with levels will be used, as we compare the tradeoff of using lower stage later.
We will only consider the time to perform the numeric factorization (i.e., $time(matrix,p)$) for a particular matrix using $p$ cores.
Fig~\ref{fig:slowdown} presents the slowdown of WSMP vs \luname , i.e., $$slowdown(matrix,p) = \frac{time(WSMP,matrix,p)}{time(Javelin,matrix,p)}.$$
In many cases, WSMP failed due to numerical constraints placed in part by the internal structure and required reordering, and we place `x' on such columns. 
Moreover, we only present up to 8 cores as WSMP does not scale past this point on either system.
Note that \luname is multiple magnitudes faster than WSMP in both serial and in parallel.
This multiple magnitude difference is due to using very light weight data structures that leads to no overhead.
WSMP (along with many other packages) uses supernode or supernode-like data structures that are too big for very sparse incomplete factorization as it does too many data movement operations per float-point operation.
As core counts increase, other packages try to reduce synchronizations by grouping like nonzero structures into these data structures, and thus reducing the serial fraction that limits strong scaling. 
However, if there does not exist many similarities in nonzero structure (as in ILU), the number of reductions is very low.
In contrast, \luname reduces the number of synchronizations by locality in memory and by using point-to-point synchronizations.

\textbf{Scalability of Javelin.}
We will define $$speedup(matrix,p) = \frac{time(matrix,1)}{time(matrix,p)}$$ for any particular system for the remainder of this  section.
We currently do not consider setup time as this is merely finding the level scheduling and copying to CSR which is done in parallel, and would be reused by other methods such as \spmv and \stri.
Moreover, almost all solver packages require some overhead to copy to structure and preorder, and we do both our copies and level-set ordering in parallel. 
\luname is $\sim10\times$ faster than WSMP in this  stage.

\begin{figure}[t]
\centering
\subfloat[14 Haswell Cores (1 Socket)]{\label{fig:hw8}\includegraphics[scale=0.28]{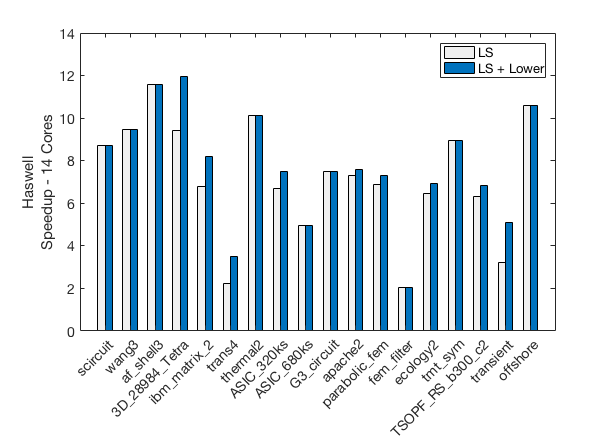}}
\\
\subfloat[28 Haswell Cores (2 Sockets)]{\label{fig:hw28}\includegraphics[scale=0.28]{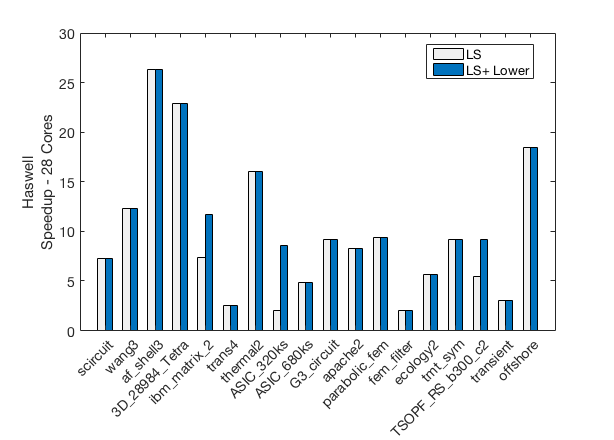}}
\caption{Speedup on Intel Haswell. The LS bars are speedup due only to level scheduling with point-to-point synchronization and the LS+Lower bars are using both level scheduling and best lower method.}
\label{fig:haswell}
\end{figure}

In Fig.~\ref{fig:haswell}, we present the speedup for \luname over all matrices on Haswell for 14 and 28 cores. 
We denote the speedup using only the level scheduling with point-to-point communications as LS, and we denote using best combination of level scheduling and lower as LS+Lower.
The geometric mean of speedup for selecting the best mixture of all methods and threads is only $9.45\times$ (not pictured), however we can see this value does not really reflect the level of speedup possible by \luname.
In general, we observe most matrices can achieve around an $8\times$ speedup using just level scheduling with point-to-point synchronizations on 14-cores.  
Several matrices are unable to reach this standard of performance.
For \texttt{ibm\_matrix\_2}, \texttt{trans4}, and \texttt{transient}, our lower method is able to boost performance closer to the level of the other matrices.
The \texttt{fem\_filter} is unable to be helped by these methods, and this is due to the numerical pattern not allowing to find many large levels, see section~\ref{sec:anal}.
As we cross socket from 14 cores to 28 cores, we notice a much larger gap in performance of difference matrices.
Matrices such as \texttt{af\_shell3} and \texttt{3D\_28984\_Tetra} are able to perform well over socket, but most cannot.
This is due to the non-uniform memory access (NUMA) between the sockets.
Currently, \luname does not account for these types of accesses.
The level scheduling depends on light weight point-to-point synchronizations using locks that are not ideal across sockets due to their required memory load.
The SR lower scheduling method does very poorly across sockets.  
This is due to having no locality in the tasks as they are taken from the tasking queue, and results in fewer matrices being boosted by using a lower method.
With the current OpenMP Scheduling, ER also has trouble with memory accesses across sockets.
ER could be improved with a more static scheduling or NUMA-aware blocking of the distribution of the lower rows. 
Despite this, no performance on 28 cores is worse than  on 14 cores, and performance of \texttt{ibm\_matrix\_2} and \texttt{TSPF\_RS\_b300\_c2} is able to be boosted by the lower methods.

\begin{figure}[tb]
\centering
\subfloat[68 KNL Cores (1 Socket)]{\label{fig:knl64}\includegraphics[scale=0.28]{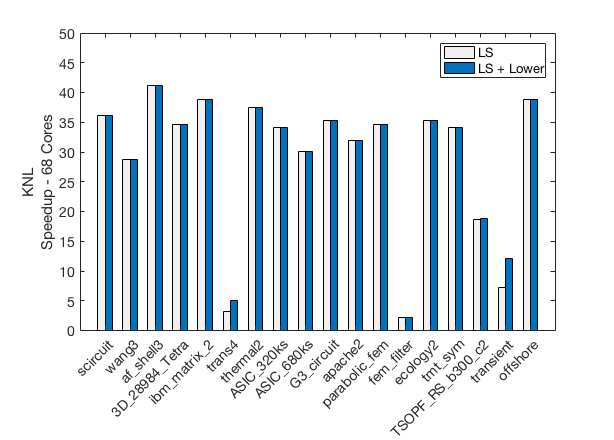}}
\\
\subfloat[68 KNL Cores X 2 Threads (1 Socket)]{\label{fig:knl128}\includegraphics[scale=0.28]{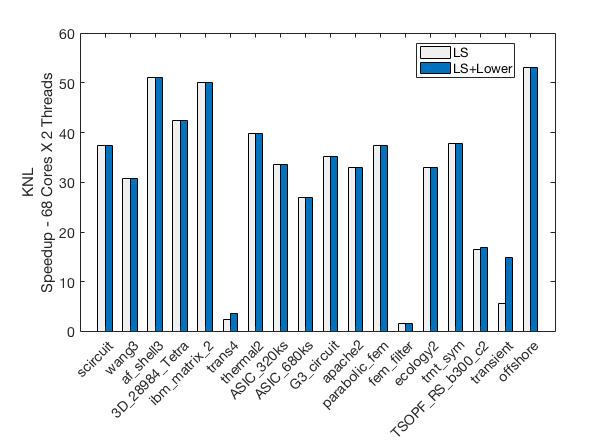}}
\caption{Speedup on Intel KNL. The LS bars are speedup only due to level scheduling with point-to-point synchronization and the LS+Lower bars are using both level scheduling and best lower method.}
\label{fig:knl}
\end{figure}

In Fig.~\ref{fig:knl}, we present the speedup performance of \luname on KNL.
KNL is a many-core system that is designed for very high levels of concurrent operations.
Each core is relatively slow compared to the fast massively out-of-order cores on Haswell.
Therefore any inefficiency in parallelizing ILU will be noticeable.
In this case, the geometric mean of the best speedup selecting the best parameters tested results in 25.1x.
Again this value is not representative of the performance we observe.
We present the speedup on 68 cores, which is all the cores in one socket with one thread in the first figure.
We see that for most matrices the speedup is around 30x using only level scheduling and point-to-point synchronizations.
As a fraction of total system use, this is less than that is achieved on Haswell.
The main reason for this is the required level of concurrency may not be able to be extracted from the matrices.
The lower methods now only help in two cases, and only by a small fraction.
This seems to be due to the tasking overhead of using an OpenMP queue with so many tasks on this number of threads as examined using Intel VTune. 
A specialized light weight tasking library is currently being constructed in \luname for this reason.
Despite this, we will observe the importance of the lower stage for \stri in the next section.
In the next figure, we test using 136 threads (i.e., 68 cores with 2 threads each) to examine the ability of using multiple threads per core.
While minor performance can be gained by some matrices, we see over-subscribing cores does not yield high results with \luname due to pressure on the memory system.
However, the performance does not generally degrade.

\vspace{-4pt}
\section{Evaluation of triangular solve}
This section provides an evaluation of sparse triangular solve (\stri) inside of \luname.
Recall that the objective of \luname was to implement a package that provides a scalable incomplete factorization that could be used as a preconditioner for an iterative solver with the structural needs to make \stri and \spmv scale.
In standard execution, the incomplete factorization may only be formed once, but \stri may be called thousands of times.
Exact iteration counts for a subset of matrices in our test suite can be found in Table~\ref{tab:iter}.
We will only consider \stri here as this is the primary call needed for methods like GMRES that use ILU, and will leave \spmv for future work that addresses preconditioners that use \spmv like successive over-relaxation.
We will compare the performance to that of a standard implementation of \stri in CSR format using OpenMP and barriers between levels in a level set ordering as done in previous works~\cite{ andrew,park}, and label it as CSR-LS.
WSMP is not reported due to its lack of performance and its inability to factor many matrices.
Using CSR-LS as the base of comparison, performance is: 
$$max speedup(m,mat,p) = \frac{time(\text{CSR-LS}, mat, 1)}{max_{i=1}^{p} \{time(m, mat, i) \}} $$
where $m$ is the method, $mat$ is the matrix, and $p$ is the number of cores.
The two methods compared from \luname are using only the level scheduling stage (LS) and using both level scheduling with the lower stage blocking that is automatically picked by \luname (LS + Lower).

\begin{figure}[tbh]
\centering
\subfloat[14 Haswell Cores (1 Socket)]{\label{fig:hsptri}\includegraphics[scale=0.24]{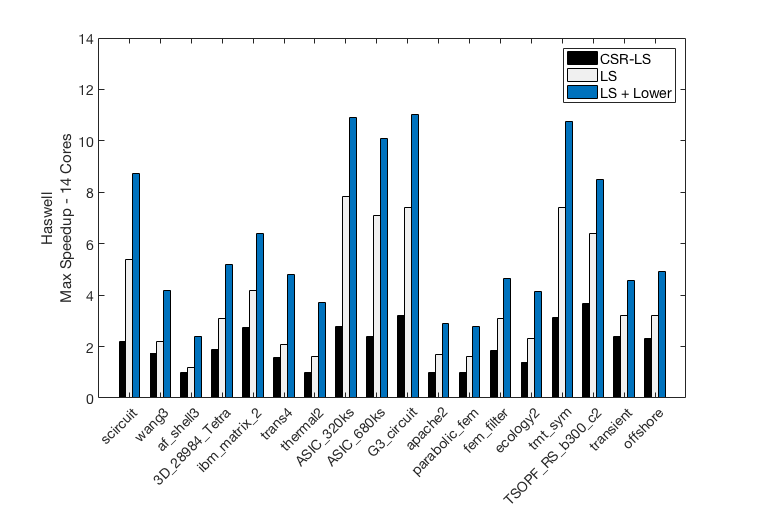}}
\\
\subfloat[68 KNL Cores (1 Socket)]{\label{fig:knlsptri}\includegraphics[scale=0.24]{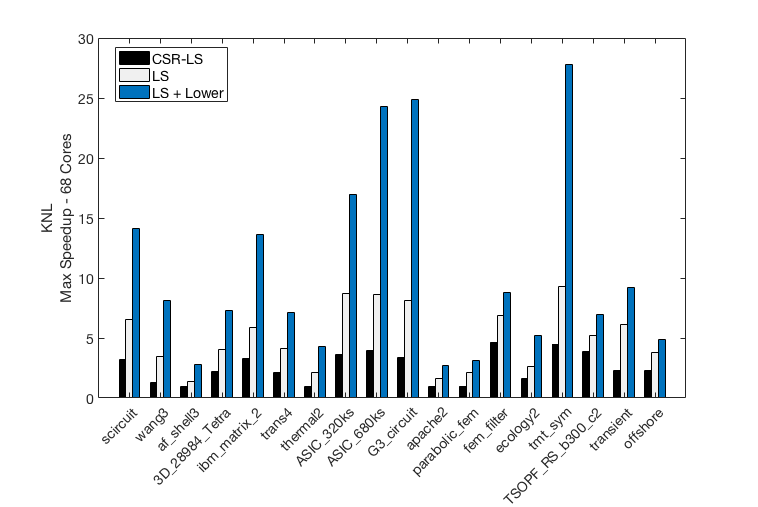}}
\caption{The maximal speedup of \stri using \luname.  The CRS-LS bars are the speedup of the standard level-set \stri implement with OpenMP.  The LS bars are due only to using the level scheduling stages of \luname, and LS + Lower is the speedup using both stages.}
\label{fig:sptri}
\end{figure}

In Fig.~\ref{fig:sptri}, we provide the maximal speedup for \stri on one socket of Intel Haswell and Intel KNL.
As is known, the traditional method of using level-sets with barriers does not scale well.
Using only the level scheduling method with pruning in \luname we are able to now have much better scaling.
However, in some cases, such as \texttt{wang3} on Haswell cores, the difference between the base case can be very small.
The lower stage blocks help performance of \stri on all matrices, and increase performance of \texttt{wang3}.
In particular, this is seen the most on Intel KNL where the lower stage does not seem to help speedup incomplete factorization as much as on Haswell cores.
We believe that this behavior is currently due to the parameters that judge between ER and SR, lower stage size, and subblocking have been tuned with more weight to \stri than ILU.
This weighting is due to an iterative method's time dominated by \stri over ILU.
Further investigation is needed to provide better balance for different matrices, but may require some estimate of conditioning of the linear system.

\vspace{-8pt}
\section{Sensitivity analysis}
\label{sec:anal}
In this section, we will examine options that can be modified in \luname, and their general effects.
We will examine the choice of ordering on iteration count and the distribution of levels for different choices of the lower stage. 

\begin{figure}[tbh]
	\centering
	\includegraphics[scale=0.26]{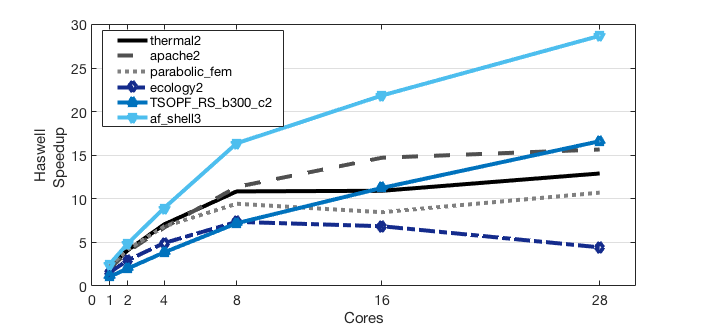}
	\caption{The relative speedup of group A matrices on Intel Haswell when the input matrix is first order with RCM. }
	\label{fig:rcm}
\end{figure}

\textbf{Iteration count.}
As previously stated, the number of iterations needed to converge to a solution will depend not only on the conditioning of the system, but the ordering of the sparse coefficient matrix.
Though some work has been done to better understand this phenomenon, it is still not fully understood why the order of the sparse coefficient matrix has such a strong impact on the convergence rate.
It is known that Coloring and ND orderings in general increase iteration count, and orders like RCM decrease iteration count.
In Table~\ref{tab:iter}, we provide the number of iterations needed to converge to a solution with relative error of $1e$-$6$ for each of the matrices in group A and using an array of different orders.
We consider SYMAMD (AMD) due to a recommendation in~\cite{order1}, Reverse Cuthill-McKee (RCM), ND, and the natural order (NAT).
Moreover, we consider the level scheduling ordering imposed on top of coefficient matrices preordered with RCM (LS-RCM) and ND (LS-ND) in order to compare how the level set ordering done in \luname will affect iteration count.
Coloring is not considered as it is known to be worse in terms of iteration than any other ordering considered here.
We note that in most cases, RCM and NAT have the fewest iterations.
However, LS-RCM is close or even better than RCM in several cases.
In Fig.~\ref{fig:rcm}, we present the speedup of group A matrices using only level scheduling with point-to-point synchronizations that are initially ordered using RCM.
The speedup is calculated with the relative base being serial with ND ordering.
We see that the speedup is comparable to those in section~\ref{sec:perf}.
Therefore, this makes either method a candidate for increased performance using \luname.
However, the speedup relative to itself is slightly less than that with ND ordering.
The largest factor in this is the size of level sets found. 
We also note that the speedup of \stri is proportional to what can be achieved by ILU.
As a result, the choice will all be up to the user to determine if the speedup will offset the number of iterations.
For example, the first 3 matrices in Fig.~\ref{fig:rcm} would most likely be best preordered with ND, since they have good speedups and few iterations.


\begin{table}[tbh]
\centering
\caption{The number of iterations need to converge based on order for group A matrices. }
\label{tab:iter}{\footnotesize
\begin{tabular}{ | c | r | r | r | r | r | r | }
\hline
\textbf{Matrix} & \textbf{AMD} & \textbf{RCM} & \textbf{ND} & \textbf{NAT} & \textbf{LS-RCM} & \textbf{LS-ND} \\ \hline
offshore & 11 & 10 & 11 & 9 & 11 & 11 \\ \hline
parabolic\_fem & 300 & 116 & 270 & 288 & 113 & 270 \\ \hline
af\_shell3 & 420 & 381 & 648 & 284 & 350 & 390 \\ \hline
thermal2 & 764 & 385 & 750 & 813 & 540 & 753 \\ \hline
ecology2 & 1532 & 889 & 1152 & 881 & 868 & 1503 \\ \hline
apache2 & 597 & 270 & 597 & 275 & 487 & 603 \\ \hline
\end{tabular}
}
\end{table}

\vspace{-6pt}
\textbf{Levels and lower size.}
Here we will provide an analysis of the number of levels selected for rows to be solved using one of the two lower methods.
In Table~\ref{tab:lvlsize}, we provide the statistics related to the level selection when considering the pattern of $lower(A+A^{T})$.
We note that there are other factors we consider in addition to the minimal row size in selecting which rows are to be permuted to the end.
These factors include the row density and the relative location. 

\begin{table}[tbh]
\centering
\caption{Level Set information of $lower(A+A^{T})$ pattern.  Lvl is the number of levels found, M is the minimal, Max is the maximum, Med is the median number of rows in a level. R-A is the number of rows moved to the end of the coefficient matrix, where A is a sensitivity parameter. }
\label{tab:lvlsize}{\footnotesize
\begin{tabular}{ | c | r | r | r | r | r | r | r | }
\hline
\textbf{Matrix} & \textbf{Lvl} & \textbf{M} & \textbf{Max} & \textbf{Med} & \textbf{R-16} & \textbf{R-24} & \textbf{R-32} \\ \hline
wang3 & 10 & 5 & 1.11e4 & 582 & 5 & 21 & 21 \\ \hline
TSOPF\_RS... & 180 & 1 & 4741 & 109 & 0 & 430 & 529 \\ \hline
3D\_28984... & 34 & 1 & 9054 & 383 & 24 & 24 & 48 \\ \hline
ibm\_mat... & 29 & 4 & 3.02e4 & 1332 & 20 & 41 & 41 \\ \hline
fem\_filter & 554 & 2 & 9003 & 3 & 1792 & 1810 & 1862 \\ \hline
trans4 & 20 & 1 & 8.57e4 & 84 & 13 & 13 & 13 \\ \hline
scircuit & 34 & 1 & 6.23e4 & 117 & 53 & 87 & 117 \\ \hline
transient & 16 & 1 & 7.82e4 & 24 & 2 & 47 & 88 \\ \hline
offshore & 74 & 1 & 3.04e4 & 2724 & 58 & 76 & 529 \\ \hline
ASIC\_320ks & 16 & 3 & 2.23e5 & 2991 & 3 & 3 & 3 \\ \hline
af\_shell3 & 630 & 1 & 2.31e4 & 5 & 1751 & 2154 & 3682 \\ \hline
parabolic\_fem & 28 & 1 & 1.02e5 & 5068 & 16 & 33 & 33 \\ \hline
ASIC\_680ks & 21 & 1 & 5.84e5& 1316 & 26 & 26 & 50 \\ \hline
apache2 & 13 & 3 & 3.22e5 & 2961 & 3 & 19 & 19 \\ \hline
tmt\_sym & 28 & 1 & 1.78e5 & 7393 & 18 & 18 & 47 \\ \hline
ecology2 & 13 & 1 & 4.42e5 & 189 & 24 & 24 & 24 \\ \hline
thermal2 & 27 & 1 & 2.72e5 & 16951 & 7 & 7 & 31 \\ \hline
G3\_circuit & 13 & 2 & 6.19e5 & 4109 & 11 & 11 & 38 \\ \hline
\end{tabular}
}
\end{table}

From Table~\ref{tab:lvlsize}, we can observe that most matrices have only tens to hundreds of levels.
Even matrices with high row density, such as \texttt{TSOPF\_RS\_b300\_c3} and \texttt{offshore}, fit into this range of levels.
We would like to see fewer levels because: 1. reduces the preprocessing time; 2. the level sets tend to be larger allowing for more concurrency.
Because of the level size distribution, the median becomes the best possible estimator of possible parallel performance.
We see that most matrices can support hundreds of concurrent threads based off this median value.
The exception to this is \texttt{fem\_filter}, \texttt{trans4}, and \texttt{af\_shell3}.
While \texttt{af\_shell3} only has a median value of 5, we discover that level scheduling still does a good job with performance as good or better than other matrices.
However, the other two do not do well with level scheduling.  
On a varying number of cores, the lower method is able to help boost \texttt{trans4}, and is not able to for \texttt{fem\_filter}.
We can observe that \texttt{trans4} has relatively few rows that ever get permuted to the end (i.e., 13). 
However, using SR, we are able to improve the speedup by a difference of $1$-$2 \times$ on socket.
We believe that we could improve selecting a better tile size and changing the tasking system to reduce overhead in the future.
Moreover, \texttt{transient} also suffers from a small median value.  
However, the lower method is able to improve the performance on socket by a factor of $\sim 2.3 \times$ on Haswell and by a factor or $\sim 1.6 \times$ on KNL, despite the current overheads in the lower method.

\begin{table}[thbp]
\caption{Level Set information of $lower(A)$ pattern. Min is the minimal and Max is the maximum number of rows in a level.}
\label{tab:nonsym}
{\footnotesize
\begin{center}
\begin{tabular}{|c|r|r|r|} \hline
\textbf{Matrix} & \textbf{Min} & \textbf{Max} & \textbf{Median} \\ \hline
TSOPF\_RS\_b330\_c2 & 1 & 4753 & 122 \\ \hline
3D\_28984\_Tetra & 4 & 9059 & 543 \\ \hline
ibm\_matrix\_2 & 1 & 30259 & 1361 \\ \hline
trans4 & 1 & 8549 & 493 \\ \hline
\end{tabular}
\end{center}
}
\end{table}

Additionally, we look at similar statistics for the $lower(A)$ pattern in Table~\ref{tab:nonsym}.
We note that we could only use this pattern if we consider just level scheduling or level scheduling with Even-Rows, but not with the Segmented-Row method.
We note that the median number of nonzeros does increase in all matrices as we would expect.
However, the increase is very small except in a few cases such as \texttt{trans4}.
Despite this increase, we could not get any real additional speedup performance for \texttt{trans4} using the $lower(A)$ pattern.
Therefore, we by default always recommend using the $lower(A+A^{T})$ pattern as it will allow for SR to be run and enable tiling for \stri.

\section{Conclusions}
\label{sec:con}
This paper presented a new scalable framework for incomplete factorization, called \luname, aimed at allowing traditional thresholding and fill-level options while being optimized for sparse matrix-vector multiplication and sparse triangular solves on current many-core systems.
This framework is designed using level scheduling of up-looking LU factorization that is implemented with light weight point-to-point synchronizations and then using a second (i.e., lower) level stage when level scheduling will no longer provide the needed degree of concurrency.
This lower stage is designed by co-optimizing the incomplete factorization and \stri after factorization.
This is achieved through reducing the serial portion and imbalance to a minimal through design, and thus helping to achieve strong scalability.  
We demonstrated that \luname could achieve good speedups on socket for both Intel Haswell and Intel KNL systems for a wide array of coefficient matrices for both ILU and triangular solve.
Moreover, we examined the effect that requiring reordering would have on the iteration count of several test systems.
This demonstration shows that, not only can we scale incomplete factorization, we can leave the system in an order that has desirable properties.



\section*{Acknowledgments}
We thank Andrew Bradley and Siva Rajamanickam for helpful discussions while at Sandia National Laboratories. 


\bibliographystyle{IEEEtran}
\bibliography{IEEEabrv,iluk}
\end{document}